\documentclass{rmaa}
\newcommand{\Msun}{\mbox{$\rm M_{\odot}$}}

\usepackage{paralist}
\usepackage{graphicx}
\title{V1898 Cygni\thanks{Based on 
  observations collected at Catania Astrophysical Observatory (Italy) and T\"{U}B\.ITAK National 
  Observatory (Antalya, Turkey).}: An interacting eclipsing binary in the vicinity of North America nebula.}

\author{
A. Dervi\c{s}o\v{g}lu,\altaffilmark{1} 
\"{O}. \c{C}ak{\i}rl{\i},\altaffilmark{1,2} 
C. \.{I}bano\v{g}lu\altaffilmark{1} and E. Sipahi\altaffilmark{1}
}

\altaffiltext{1}{Ege University, Science Faculty, Department of Astronomy and Space Sciences, 35100 Bornova, \.{I}zmir, TURKEY}
\altaffiltext{2}{T{\"U}B{\.I}TAK National Observatory, Akdeniz University Campus, 07058 Antalya, TURKEY}
\shortauthor{Dervi\c{s}o\v{g}lu et al.}
\shorttitle{V1898 Cygni}
\fulladdresses{
\item Ege University, Science Faculty, Astronomy and Space Sciences Dept., 35100 
Bornova, \.{I}zmir, Turkey. $e-mail$: ahmetdervisoglu@mail.ege.edu.tr}

\listofauthors{A. Dervi\c{s}o\v{g}lu, 
\"{O}. \c{C}ak{\i}rl{\i},
C. \.{I}bano\v{g}lu and E. Sipahi}
\indexauthor{Dervi{\c s}o{\v g}lu, A.}
\indexauthor{\c{C}ak{\i}rl{\i}, \"{O}}
\indexauthor{\.{I}bano\v{g}lu, C.}
\indexauthor{Sipahi, E.}

\abstract{
We present spectroscopic observations of the double-lined Algol type eclipsing binary V1898\,Cyg. 
Analyses of the BV light curves and RVs led to determination of the fundamental stellar parameters of the V1898\,Cyg's components.   The absolute parameters for the stars are derived as: M$_1$=6.054$\pm$0.037  
M$_{\odot}$, M$_2$=1.162$\pm$0.011 M$_{\odot}$, R$_1$=3.526$\pm$0.009 R$_{\odot}$, R$_2$=2.640$\pm$0.010 
R$_{\odot}$, T$_{eff_1}$=18\,000$\pm$600 K, and T$_{eff_2}$=6\,200$\pm$200 K.
The residuals between the observed and computed times of mid-eclipses were analysed and a rate of the period change 
$\dot{P}/P= 6.68 \times 10^{-7}\,yr^{-1}$ was obtained and a mass transfer rate of $1.88\times10^{-7}\,$ \Msun  in a year is estimated.We have calculated the distance to the system of V1898\,Cyg as 501$\pm$5 pc using the infrared JHK magnitudes and bolometric corrections for the primary star. The components of the system's proper motions present some indications about membership of the North America nebula. 
}
\resumen{}
\addkeyword{ stars: binaries: close}
\addkeyword{ binaries: eclipsing}
\addkeyword{ binaries: general}
\addkeyword{ binaries: spectroscopic}
\addkeyword{ stars: individual: V1898\,Cyg}
\begin{document}
\maketitle

\section{Introduction}
\label{}
V1898\,Cyg (HD 200776, BD +45$^{\circ}$ 3384, 2MASS J21035377+4619499, HIP 103968, V=7$^m$.81, (B-V)=+0$^m$.01) was discovered 
to be single-lined spectroscopic binary by Abt et al. (1972).  HD 200776 was included in the list of bright OB stars to be 
observed for determination of galactic rotational constants and other galactic parameters. Their spectroscopic observations 
yield that HD 200776 is a spectroscopic binary with an orbital period of 2.9258 days. They also calculated the preliminary 
elements and mass function for the system. McCrosky and Whitney (1982) searched for photometric variations some short-period 
spectroscopic binaries including HD 200776 given in the Seventh Catalogue of the Orbital Elements of Spectroscopic Binary 
System (Batten et al. 1978). They observed abrupt drops, with an amount of 0$^m$.2-0$^m$.4, in the brightness of the system 
which were inconsistent with the orbital period proposed by Abt et al (1972). Photometric observations in B and V-bandpass 
made by Halbedel (1985) revealed that HD 200776 is an eclipsing binary with both eclipses are nearly identical, in contrast 
to the spectroscopic observations. He proposed a new orbital period of 3.0239 days, 3 per cent longer than that given by Abt 
et al (1972). Caton and Smith (hereafter CS, 2005) published new light curve and times of mid-eclipses as well as a new orbital 
period of 1.5131273 days, nearly half that of given by Halbedel (1985). Soon later Dallaporta and Munari (hereafter DM, 2006) 
presented complete and accurate BV light curves of HD 200776 as well as three times for mid-primary eclipse. Fortunately the 
same comparison star was used in the photometric observations of Halbedel (1985) and DM. No further spectroscopic observations 
were made into this eclipsing-spectroscopic binary after Abt et al. (1972).   

The main aims of this study are: (1)to detect some lines of the secondary component;(2)to reveal radial velocities for both 
components; (3) to solve the radial velocity curves for the primary and secondary components on the basis of new observations 
in order to reveal accurate masses and radii; and (4)to determine the rotational velocities of the components and compare with 
orbital synchronization.

\section{Spectroscopic Observations}  
The spectra were obtained from several telescopes during the course of three years, beginning the year of 2007. Table\,1 lists the full set of observations. The 
first set was observed with the \'{e}chelle spectrograph (FRESCO) at the 91-cm telescope of Catania Astrophysical Observatory. Spectroscopic 
observations have been performed with the spectrograph is fed by the telescope through an optical fibre ($UV$--$NIR$, 100 $\mu$m core 
diameter) and is located, in a stable position, in the room below the dome level. Spectra were recorded on a CCD camera equipped with 
a thinned back--illuminated SITe CCD of 1k$\times$1k pixels (size 24$\times$24 $\mu$m). The cross-dispersed \'{e}chelle configuration 
yields a resolution of about 22\,000, as deduced from the full width at half maximum of the lines of the Th--Ar calibration lamp. The 
spectra cover the wavelength range from 4300 to 6650 {\AA}, split into 19 orders. In this spectral region, and in particular in the blue 
portion of the spectrum, there are several lines useful for the measure of radial velocity, as well as for spectral classification of 
the stars.

The system was also observed with the Turkish Faint Object Spectrograph Camera (TFOSC) attached to the 1.5 m RTT150 telescope on 
August 07-20, 2010 under good seeing conditions. Further details on the telescope and the spectrograph can be found at 
http://www.tug.tubitak.gov.tr. The wavelength coverage of each spectrum was 4100-9000 {\AA} in 11 orders, with a resolving 
power of $\lambda$/$\delta\lambda$\, 7\,000 at 6563 {\AA}.

The electronic bias was removed from all spectra and we used the {\sc crreject} task of IRAF\footnote{IRAF is distributed by the National Optical Observatory, which is operated by the Association of the Universities 
for Research in Astronomy, inc. (AURA) under cooperative agreement with the National Science Foundation} for cosmic ray removal. The \'{e}chelle spectra 
were extracted and wavelength calibrated by using a Fe-Ar and Th-Ar lamp source with help of the IRAF echelle package. The stability 
of the instruments were checked by cross correlating the spectra of the standard star against each other using the {\sc fxcor} task in 
IRAF. The standard deviation 
of the differences between the velocities measured using {\sc fxcor} and the velocities in Nidever et al. (2002) was about 1.1 km s$^{-1}$.

Twenty-eight spectra of V1898\,Cyg were collected during the two different seasons. Typical exposure times for the V1898\,Cyg 
spectroscopic observations were between 2400 and 2600\,s for Catania telescope and 1200\, s for RTT150 telescope. The signal-to-noise
ratio ($S/N$) achieved was between 70 and 115, and $\sim$150 depending on atmospheric condition. $\alpha$ Lyr (A0V), 59 Her
(A3IV), $\iota$ Psc (F7V), HD 27962 (A2IV), and $\tau$ Her (B5IV) were observed during each run as radial velocity and/or 
rotational velocity templates. The average $S/N$ at continuum in the spectral region of interest was 150--200 for the 
standard stars.

\begin{figure}
\begin{center}
\includegraphics[width=12cm]{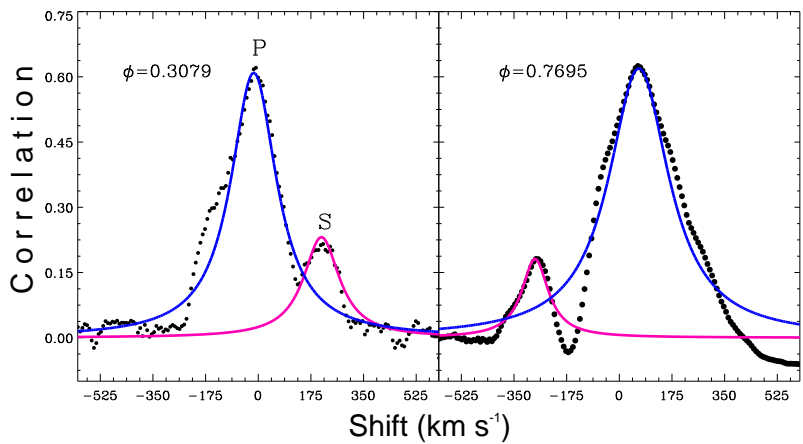} 
\end{center}\caption{Sample of Cross Correlation Functions (CCFs) between V1898\,Cyg and the radial velocity template 
spectrum (Vega) at four different orbital phases.}
\label{ccf;fig1}
\end{figure}

\begin{figure}
\begin{center}
\includegraphics[width=9cm]{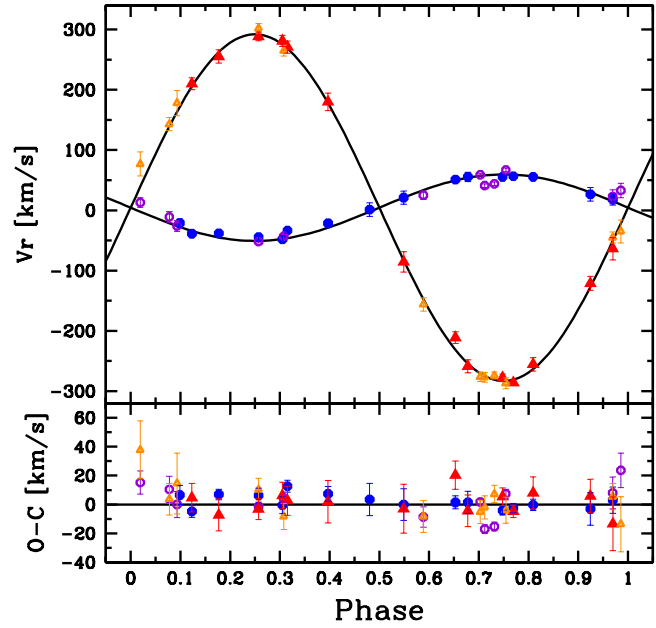}
\end{center}\caption{Radial velocities of the primary (dots) and secondary stars (triangles) are folded on an orbital period of 1.513126 days. The 
velocities obtained at the Catania observatory are indicated by filled symbols while those obtained at the National Observatory of Turkey by open 
symbols. The vertical lines show error bars of each radial velocity. The residuals between the observed and computed RVs are plotted in the lower panel. 
}
\label{RV;fig2}
\end{figure}

\begin{table}
\caption{Radial velocities of the V1898\,Cyg's components. The columns give the heliocentric Julian date, the
orbital phase, the radial velocities of the two components with the corresponding errors and residuals. }
\begin{center}
\begin{tabular}{@{}lcrrrrrrr@{}}
\hline
\textsf {HJD} & Phase& \multicolumn{3}{c}{Star 1 }& \multicolumn{3}{c}{Star 2 } &  Remarks \\
  2\,400\,000+ &  & \textsf{{\bf V$_p$}} & $\sigma$ & O-C & \textsf{{\bf V$_s$}} & $\sigma$& O-C & \\
\hline
54327.55412	&0.5489	&21.1	&10.9 &0.0     &-85.5  &16.9  & -3.0   &a\\
54328.50480	&0.1772	&-38.1	&3.6  & 6.9    &255.3  &11.1  & -7.2   &a\\
54329.46135	&0.8093	&55.5	&3.9  &-0.3    &-255.5 &11.1  &  8.0   &a\\
54330.47685	&0.4805	&1.2	&11.1 & 3.5    &...    &...   & ...    &a\\
54331.44914	&0.1231	&-38.8	&4.2  &-4.7    &210.1  &9.9   &  4.7   &a\\
54335.45358	&0.7695	&56.7	&3.3  &-2.5    &-285.8 &4.3   & -4.8   &a\\
54336.40281	&0.3969	&-21.5	&5.1  & 7.3    &179.9  &14.6  &  1.9   &a\\
54337.46485	&0.0988	&-21.1	&6.6  & 6.5    &...    &...   & ...    &a\\
54338.44704	&0.7479	&55.5	&3.1  &-4.1    &-277.7 &6.1   &  5.5   &a\\
54360.40110	&0.2571	&-44.4	&5.1  & 6.3    &288.6  &7.2   & -3.2   &a\\
54361.41090	&0.9245	&16.6	&11.2 &-3.0    &-121.1 &11.6  &  5.8   &a\\
54362.55110	&0.6780	&55.5	&7.7  & 1.4    &-258.6 &10.9  & -4.3   &a\\
54363.50010	&0.3052	&-47.9	&5.9  &-0.5    &281.1  &9.2   &  6.1   &a\\
54364.50527	&0.9695	&18.0	&9.1  & 3.1    &-63.7  &18.5  & -13.4  &a\\
54365.53939	&0.6530	&51.1	&4.5  & 1.4    &-211.1 &9.8   &  20.3  &a\\
54366.54137	&0.3152	&-33.5	&4.1  & 12.6   &270.9  &10.1  &  2.6   &a\\
55387.53130	&0.0777	&-11.0	&9.0  & 10.4   &143.0  &11.0  &  3.6   &b\\
55390.46920	&0.0193	&13.0	&8.0  & 15.2   &77.0   &20.0  &  37.8  &b\\
55390.58090	&0.0932	&-26.0	&9.0  & 0.0    &178.0  &21.0  &  14.6  &b\\
55391.50380	&0.7031	&59.0	&2.0  & 1.8    &-276.0 &8.0   & -5.2   &b\\
55391.58150	&0.7544	&67.0	&3.0  & 7.4    &-287.0 &9.0   & -3.9   &b\\
55392.41900	&0.3079	&-43.0	&3.0  & 4.1    &265.0  &9.0   & -8.3   &b\\
55393.41980	&0.9694	&23.0	&11.0 & 8.0    &-45.0  &9.0   &  5.5   &b\\
55394.57290	&0.7314	&44.0	&3.0  &-15.2   &-274.0 &6.0   & 7.2    &b\\
55396.47080	&0.9857	&33.0	&12.0 & 23.6   &-35.0  &19.0  & -13.6  &b\\
55397.38320	&0.5887	&25.0	&7.0  &-8.6    &-156.0 &11.0  & -8.3   &b\\
55397.56950	&0.7119	&41.0	&3.0  &-17.0   &-277.0 &8.0   & -2.0   &b\\
55398.39480	&0.2573	&-52.0	&2.0  &-1.4    &302.0  &8.0   & 10.2   &b\\

\hline
\end{tabular}
\end{center}

\begin{list}{}{}
\item[Remarks:]{\small  (a) Based on Catania and (b) on TUG observations. }

\end{list}
\end{table}

\section{Spectroscopic analysis}
Double-lined spectroscopic binaries reveal two peaks, displacing back and forth, in the cross-correlation 
function (CCF)  between variable and the radial velocity template spectrum as seen in Fig.~1. The location of 
the peaks allows to measure of the radial velocity of each component at the time of observation. The cross-correlation 
technique applied to digitized spectra is now one of the standard tools for the measurement of radial velocities 
in close binary systems.

The radial velocities of V1898\,Cyg were obtained by cross--correlating of \'{e}chelle orders of 
V1898\,Cyg spectra with the spectra of the bright radial velocity standard stars $\alpha$ Lyr (A0V), 59 Her (A3IV) and 
$\iota$ Psc (F7V) (Nordstr\"om et al., 2004). For this purpose the IRAF task {\sc fxcor} was used. 

Fig. 1 shows examples of CCFs of V1898\,Cyg near the first and second quadrature. The two non-blended peaks correspond 
to each component of V1898\,Cyg. We applied the cross-correlation technique to five wavelength regions with well-defined 
absorption lines of the primary and secondary components. These regions include the following lines: Si\,{\sc iii} 
4568 \AA, Mg\,{\sc ii} 4481 \AA, He\,{\sc i} 5016 \AA, He\,{\sc i} 4917 \AA, He\,{\sc i} 5876 \AA. The stronger CCF 
peak corresponds to the more massive component that also has a larger contribution to the observed spectrum. To better 
evaluate the centroids of the peaks (i.e. the radial velocity difference between the target and the template), we 
adopted two separate Gaussian fits for the case of significant peak separation.

The radial velocity measurements, listed in Table 1 together with their standard errors, are weighted means of
the individual values deduced from each order. The observational points and their error bars are displayed in Fig. 2 as a 
function of orbital phase as calculated by means of the linear part of the ephemeris given in Eq.\,2. The radial 
velocities of the secondary component of V1898\,Cyg are presented for the first time in this study. The simultaneous 
analysis both curves gives the semi-amplitude of the more massive, more luminous component to be K$_1$=55.2$\pm$0.8 
km s$^{-1}$ and K$_2$=287.6$\pm$2.1 km s$^{-1}$ for the secondary component with a systemic velocity of $4.4\pm$0.8 km s$^{-1}$.

\subsection{Spectral classification}
The spectral types of the stars can be found either photometry or spectroscopy and/or both. The apparent visual magnitude and 
colours of V1898\,Cyg were estimated by Hiltner (1956) as V=7$^m$.81, B-V=0$^m$.01, U-B=-0$^m$.82. However the apparent magnitudes 
are given by Reed (2003) as 7$^m$.0, 7$^m$.80 and 7$^m$.82 in the U, B and V passbands, respectively. On the other hand, The B-V 
colour of the system at outside of eclipse was determined as 0$^m$.01 by Halbedel (1985). We computed the B-V colour of the system 
at the maxima as 0$^m$.036 using the data given by Dallaporta and Munari (2006). The combined spectral types are given as B1 IVp by 
Abt et al. (1972) and B2 III by Kennedy and Buscombe (1974). The infrared magnitudes of the system are given by Cutri et al. (2003) 
as J=7$^m$.697, H=7$^m$.718 and K=7$^m$.757. Unfortunately, intermediate- and narrow-band photometric measurements are not available 
for the system. Bessell et al.(1998) derive reddening-independent Q-parameter from theoretical colours as Q=(U-B)-0.71 (B-V). They 
also predict interstellar reddening for the main-sequence OB stars as E(B-V)=(B-V)-((U-B)-0.71(B-V))/3. Using the observed (U-B) 
and (B-V) colours by Hiltner (1956) we find E(B-V)=0$^m$.28, while for Dallaporta and Munari (2006) estimate the reddening as 0$^m$.31.  

We computed the intrinsic colours of the primary component using the JHK magnitudes as $J-H=-0^m.021\pm0^m.044$ and $H-K=-0^m.039\pm0^m.033$. V1898\,Cyg 
locates between supergiants and dwarfs in the infrared (J-H)-(H-K)diagram (Tokunaga 2000). Using the the equations given by Strai{\v z}ys et al. 
(2008) we estimated interstellar reddening as $E(J-H)=0^m.106\pm0^m.060$ and  $E(H-K)=0^m.052\pm0^m.060$. Using the transformation equation 
given by Bessell et al.(1998)we find  $E(B-V)=0^m.286$, in a very good agreement with that found by the UBV colours. Since the observed 
common colour index is 0$^m$.036 one obtains an intrinsic B-V colour of -0$^m$.25  which corresponds to a B1V star in the calibrations 
of Papaj et al. (1993).        
    
We have used our spectra to determine the spectral type of the primary component of V1898\,Cyg. We have followed the procedures of 
Hern\'andez et al. (2004), choosing helium lines in the blue-wavelength region, where the contribution of the secondary component to 
the observed spectrum is almost negligible. From several spectra we measured  $EWs$ of $\rm He I\lambda 4026, 4144, 4387, 4922$ 
as $0.867\pm 0.044, 0554\pm0.062, 0.497\pm 0.090, 0.768\pm0.028$\,\AA, respectively. Then we used the $EW$-spectral type  diagrams given 
by Hern\'andez et al. (2004). The $EWs$ of the helium lines indicate that the spectral type of the primary component is $B1.8\pm0.6$ which 
is in a agreement with that obtained from infrared photometry. The calibration of Papaj et al. (1993) gives an effective temperature of 
18700 K and B-V of -0$^m$.21 for a B2V star, 16800 K and -0$^m$.19 for the same spectral type but for a giant star. Therefore, we estimated 
an effective temperature of $18000\pm600$ K for the primary component of V1898\,Cyg.    

\subsection{Reddening}
The measurement of reddening is a key step in determining the distance of stars. V1898\,Cyg locates in the direction of the NAP, in which 
reddening varies from one place to other. We estimated the reddening in the B-V colour as 0$^{m}$.29 using the infrared colours. On the 
other hand we find  $E(B-V)=0^{m}$.25 for a star of B2V and $E(B-V)=0^{m}$.23 for a star of B2III. The photometric and spectroscopic 
determinations of the interstellar reddening seem to be in a good agreement within $3\,\sigma$ error. Our spectra cover the interstellar 
Na{\sc i} (5890 and 5896 \AA) doublet, which is excellent estimator of the reddening as demonstrated by Munari \& Zwitter (1997). They 
calibrated a tight relation linking the Na {\sc i} D1 (5890 \AA) equivalent widths with the E(B-V) reddening. On spectra obtained at 
quadratures, lines from both components are un-blended with the interstellar ones, which can therefore be accurately measured. We 
derive an equivalent width of 0.52$\pm$0.06 \AA~ for Na{\sc i} D1 line, which corresponds to E(B-V)= 0$^{m}$.33$\pm$0.09. Since the 
star locates nearly at the galactic plane($l$=87\arcdeg .60, $b=$-0\arcdeg .34) and near the edge of North America nebula (NAN) 
such a reddening in the optical wavelengths is expected.

\subsection{Rotational velocity}
The width of the cross-correlation profile is a good tool for the measurement of $v \sin i$ (see, e.g., 
Queloz et al. 1998). The rotational velocities ($v \sin i$) of the two components were obtained by 
measuring the FWHM of the CCF peaks in nine high-S/N spectra of V1898\,Cyg acquired close to the 
quadratures, where the spectral lines have the largest Doppler-shifts. In order to construct a 
calibration curve FWHM--$v \sin i$, we have used an average spectrum of HD~27962, acquired with 
the same instrumentation. Since the rotational velocity of HD~27962 is very low but not zero 
($v \sin i$ $\simeq$11 km s$^{-1}$, e.g., Royer et al. (2002) and references therein), it could be 
considered as a useful template for A-type stars rotating faster than $v \sin i$ $\simeq$ 10 
km s$^{-1}$. The spectrum of HD~27962 was synthetically broadened by convolution with rotational 
profiles of increasing $v \sin i$ in steps of 5 km s$^{-1}$ and the cross-correlation with the original 
one was performed at each step. The FWHM of the CCF peak was measured and the FWHM-$v \sin i$ 
calibration was established. The $v \sin i$ values of the two components of V1898\,Cyg were derived 
from the FWHM of their CCF peak and the aforementioned calibration relations, for a few wavelength 
regions and for the best spectra. This gave values of 110$\pm$5 km s$^{-1}$ for the primary star 
and 90$\pm$9 km s$^{-1}$ for the secondary star.

\begin{table}
\caption{Times of mid-eclipses for V1898\,Cyg and the O-C residuals (see text)}
\label{O-C values}
\begin{center}
\begin{tabular}{cccccr}
\hline
Minimum time	& Epoch	& O-C(I) &O-C(II) &O-C(III)	& Ref.\\
(HJD-2\,400\,000) & & & & \\
 \hline
45960.6758	&-3126	&-0.0371	&0.0139	&0.0004	&1\\
45963.6986	&-3124	&-0.0406	&0.0105	&-0.0030	&1\\
46010.6101	&-3093	&-0.0355	&0.0151	&0.0018	&1\\
46013.6351	&-3091	&-0.0367	&0.0138	&0.0006	&1\\
50690.6948	&0	&0.0000	&0.0010	&0.0010	&3\\
52169.7772	&977.5	&0.0174	&0.0027	&0.0014	&3\\
52185.6636	&988	&0.0161	&0.0013	&0.0000	&3\\
52895.3220	&1457	&0.0259	&0.0036	&0.0007	&2\\
52901.3740	&1461	&0.0255	&0.0031	&0.0002	&2\\
52928.6107	&1479	&0.0262	&0.0035	&0.0005	&3\\
53207.7802	&1663.5	&0.0269	&0.0013	&-0.0025	&3\\
53226.6966	&1676	&0.0294	&0.0036	&-0.0003	&3\\
53246.3663	&1689	&0.0287	&0.0027	&-0.0013	&2\\
53270.5757	&1705	&0.0284	&0.0021	&-0.0020	&3\\
54443.2559	&2480	&0.0483	&0.0096	&0.0011	&4\\
54443.2565	&2480	&0.0489	&0.0102	&0.0017	&4\\
54691.4070	&2644	&0.0494	&0.0080	&-0.0016	&4\\
54691.4089	&2644	&0.0513	&0.0099	&0.0003	&4\\
54691.4097	&2644	&0.0521	&0.0107	&0.0011	&4\\

\hline
\end{tabular}
\end{center}
\begin{list}{}{}
\item[Ref:]{\small (1) Halbedel (1985), (2) Dallaporta and Munari (2006), (3) Caton and Smith (2005), (4) Br{\'a}t et al. (2008)}
\end{list}
\end{table}

\section{Times of minima and the orbital period}
Times of mid-eclipses were published by Halbedal(1985), CS, DM and Br{\'a}t 
et al. (2008). These times of eclipses were presented in Table 2. The O-C(I) residuals are computed using the light elements 
given below,
\begin{equation}
Min I(HJD)=2\,450\,690.6948+1^d.51311 \times E
\end{equation}
where the orbital period is adopted from DM. The behaviour of the deviations from the linear light 
elements O-C(II) with respect to the epoch numbers suggests an upward curved parabola. Therefore, a parabolic fit to the data 
gives,
\begin{equation}
Min I(HJD)=2\,450\,690.6938(8)+1^d.5131260(2) \times E+1.38(13) 10^{-9} \times E^2.
\end{equation}
     
The standard mean errors in the last digits are given in the parentheses. The coefficient of the quadratic term is positive which 
indicates that the orbital period of the system is increasing with the epoch number. Such a quadratic ephemeris appears to a very 
good representation of the orbital period change of V1898\,Cyg, as well as other interacting Algols. This quadratic behaviour of 
the O-C(II) residuals, plotted in the upper panel of Fig.3, is an indication of the secular period increase for the system. In the 
bottom panel of Fig.3 the O-C(III) residuals with respect to ephemeris (2)  are plotted, which illustrates a good agreement between 
the timings and new ephemeris. It is known that the classical Algols have an evolved less massive component which fills its Roche 
lobe and transfers its mass to the more massive primary star through Lagrangian L$_1$ point of the system.  The orbital period of 
the system is increasing at the average rate of  $\dot{P}/P= 6.68 (\pm 0.63)\times 10^{-7}\,yr^{-1}$ which means that the orbital 
period was increased by about 0.38($\pm$0.04) seconds in the last 24 years. The sum of squares of residuals for the parabolic fit 
is $3.672\times10^{-5}\,$ d$^2$. We limited times of mid-eclipses covering about 24 yrs. The time span of the observations is very short to reveal any abrubt period change caused by the fast mass transfer  phenomenon. Of course the
 observations will be obtained in the coming years indicate some hints about the nature of orbital period
  change. Assuming a conservative mass transfer from the less massive component to the more massive primary 
star we estimate an amount of $ 1.88 (\pm 0.17)\times 10^{-7}$ \Msun in a year. 

\begin{figure}
\begin{center}
\includegraphics[width=7cm]{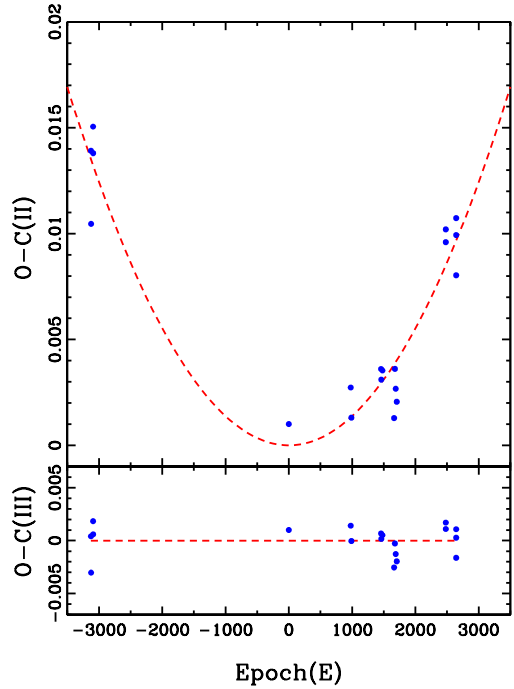} 
\end{center}\caption{The O-C(II) residuals plotted versus the epoch number for V1898\,Cyg . A least-squares quadratic fit to the 
residuals was shown by dashed line (upper panel). In the bottom panel the O-C(III)residuals, the deviations from quadratic 
fit, are also plotted.}
\label{ccf;fig.3}
\end{figure}

\begin{figure}
\begin{center}
\includegraphics[width=7cm]{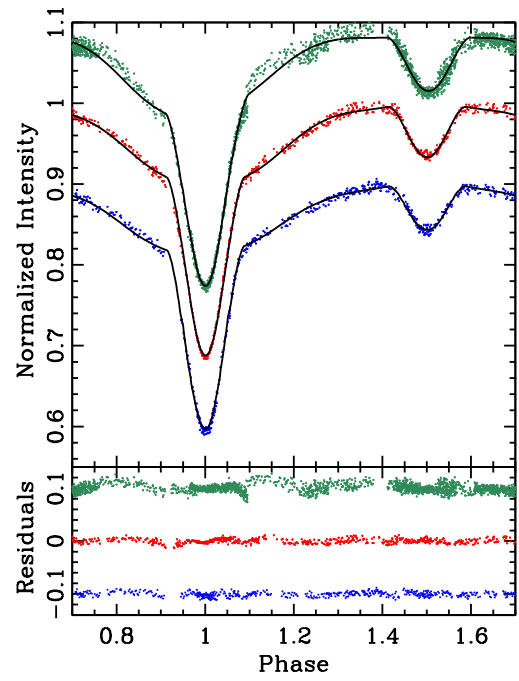}
\end{center}\caption{Comparison of the observed and computed light curves of V1898\,Cyg. From top to bottom the CS-V, DM-B and DM-V light curves, respectively. In the lower panel residuals of the fit have been plotted to show the goodness of the fit.}

\label{LC;fig5}
\end{figure}

\section{Analysis of the light curves}
Three light curves V1898\,Cyg were obtained and the observational data were published. The first photometric observations  obtained by 
Halbedel (1985) between July and November, 1985. The second photometric observations made by CS from August 22, 2001 to October 26, 
2004. The star was observed by DM from July 22, 2003 to September 17, 2004. Only the V-band light curve and observational data of 
CS were published. However,the B and V light curves of DM were published. The V light curves obtained in two studies are asymmetric 
in shape, furthermore differ from each other. The brightness of the system shows a fast decrease from phase 0.75 up to external 
contact. However,the increase in brightness following the primary eclipse is not steeper.The distortion of the light curve preceding 
the primary eclipse is larger in the CS light curve. A remarkable change in the light curve is seen at the phase interval between 
0.09 and 0.42. The total brightness of the system in the V-bandpass at this phase interval is greater by about 0$^m$.03 in the CS 
light curve than that in DM's. However, CS report that the brightness of their primary comparison star showed light variations 
during the observations. Moreover the depth of the secondary eclipse is larger by about 0$^m$.007 in the light curve of CS than 
that of DM. We should note that there is also a slight asymmetry in both the DM's B- and V-passband light curves. This feature 
is attributed to the transferring material from cool secondary to the high temperature primary star which occults a small area 
of the primary star just before the deeper eclipse.    

Acerbi and Barani (2007) analysed the DM's light curves. Since the spectroscopic mass-ratio was not available at that time they 
started to the analysis by deriving the photometric mass-ratio.Their preliminary analysis indicated that the mass-ratio for the 
system was about 0.30. Assuming an effective temperature of 20\,183 K for the primary component and the secondary  less massive 
star fills its corresponding Roche-lobe, i.e. semi-detached configuration, they arrived at a preliminary elements for the 
system. Their orbital parameters were: $i=70^{\circ}$, $r_1=0.3196$ and $r_2$=0.2795 and T$_2$=7\,500 K. Since the light curves 
were asymmetric they reported that the agreement between the computed and observed light curves were not very satisfactory, 
the sum of squares of residuals was about 0.814.             

In order to analyse the light curves  we choose the Wilson--Devinney (W--D) code implemented into the PHOEBE package tool by 
Prsa \& Zwitter (2005). Preliminary analysis indicates that the system is a classical Algol, the secondary component fills its 
corresponding Roche lobe. Therefore Mode--5 is applied. WD code is based on Roche geometry which is sensitive to the mass ratio 
which is taken from RV analysis as 0.192$\pm$0.002. Gravity--darkening exponents $g_1$ = 1, $g_2$ = 0.32 and bolometric 
albedos $Alb_{1}$=1, $Alb_{2}$=0.5 were set, i.e. the more massive star has a radiative envelope while the less massive 
secondary convective atmosphere. We used the non-linear square-root limb--darkening and the bolometric limb-darkening 
coefficients from the tables Diaz-Cordoves, Claret, and Gimenez (1995) and van Hamme(1993).    
 
The orbital inclination ($i$), effective temperature of the secondary star (T$_2$), surface potential of the primary 
($\Omega_1$), phase shift ($\Delta \phi$), and fractional luminosity of the primary ($L_1$) were taken as adjustable 
parameters. The other parameters were fixed. The iterations were carried out automatically until convergence and a 
solution was defined as the set of parameters for which the differential corrections were smaller than the probable 
errors. The final results obtained by separate analysis of three light curves are listed in Table 3 and the computed 
light curves are shown as continuous lines in Fig. 4. The uncertainties assigned to the adjusted parameters are the
internal errors provided directly by the WD code. As seen in Table 3 the sum of residuals-squares are 0.0109 and 
0.0123 for the B- and V-bandpass light curves, indicating 75-times smaller than the analysis made by Acerbi and 
Barani (2007) of the same data. In the bottom panel of Fig.4 the residuals between observed and computed intensities 
are also plotted. The residuals reveal that the binary model may represent the observed DM's light curves 
successfully. However, the computed light curve differs mostly from fourth contact to beginning of secondary 
eclipse in the CS light curve.

\begin{table}
  \caption{Results of individual light curve analyses for V1898\,Cyg.}
  \label{parameters}
  \begin{center}
  \begin{tabular}{lccr}
  \hline
   Parameter & CS V & DM B & DM V\\
   \hline
   $i (^{\circ})$ 				& 74.20$\pm$0.02  & 73.05$\pm$0.03	 & 73.03$\pm$0.03		\\
%   $e (^{\circ})$ 				& 0.127$\pm$0.007  	\\
%   $\omega (^{\circ})$ 			& 155$\pm$4  			\\
   $T_{1}$ (K) 					& \multicolumn{3}{c}{18\,000[\textsf{Fix}]}   		\\
   $T_{2}$ (K)					& 6582$\pm$65 & 6205$\pm$76 & 6109$\pm$50 		\\
   $\Omega_{1}$ 				& 3.0887$\pm$0.0085 & 3.2885$\pm$0.0116 & 3.2784$\pm$0.0103 \\
   $\Omega_{2}$ 				& & 2.2122[\textsf{Fix}]  &  	\\
   $q_{spec}$ 					& &0.1918[\textsf{Fix}]   & 					\\
   $L_{1}/(L_{1+2})$ 			& 0.9563$\pm$0.0013 & 0.9801$\pm$0.0017 & 0.9490$\pm$0.0016	\\
   $r_1$						& 0.3513$\pm$0.0013 & 0.3296$\pm$0.0013 & 0.3286$\pm$0.0012	\\
   $r_2$						& 0.2464$\pm$0.0012 & 0.2464$\pm$0.0012 & 0.2464$\pm$0.0012 \\
   $\Delta \phi$				& 0.0022$\pm$0.0001 & 0.0008$\pm$0.0001	& 0.0008$\pm$0.0001	\\
   $\Sigma(res)^2$				& 0.1196 & 0.0109 & 0.0123  					\\
  \hline
  \end{tabular}
  \end{center}
  \begin{list}{}{}
\item[Ref:]{\small $r_1,r_2$: Relative volume radii, CS V: Caton and Smith's (2005) V-band light curve, DM B and 
DM V: The B- and V-band light curves of Dallaporta and Munari (2006). The errors quoted for the adjustable parameters 
are the formal errors determined by the WD-code.}
\end{list}

\end{table}

\begin{table}
 \setlength{\tabcolsep}{2.5pt} 
  \caption{Fundamental parameters of V1898\,Cyg.}
  \label{parameters}
  \begin{center}
  \begin{tabular}{lcc}
  \hline
%  & \multicolumn{2}{c}{V1898\,Cyg} 		\\
   Parameter 						& Primary	&	Secondary				\\
   \hline
   Mass (M$_{\odot}$) 				& 6.054$\pm$0.037 & 1.162$\pm$0.011		\\
   Radius (R$_{\odot}$) 			&3.526$\pm$0.009 & 2.640$\pm$0.010		\\
   $T_{eff}$ (K)					& 18 000$\pm$ 600	& 6 200$\pm$200      	\\
   $\log~(L/L_{\odot})$				& 3.071$\pm$0.029	& 0.957$\pm$0.085       	\\
   $\log~g$ ($cgs$) 				& 4.125$\pm$0.002 & 3.660$\pm$0.004		\\
   Spectral Type					& B2IV$\pm$1  	& G2III$\pm$1    			\\
   $a$ (R$_{\odot}$)				&\multicolumn{2}{c}{10.714$\pm$0.022}	\\
   $i$ ($^{\circ}$)					&\multicolumn{2}{c}{73.05$\pm$0.03} 		\\
   $d$ (pc)							& \multicolumn{2}{c}{501$\pm$5}			\\
   $(vsin~i)_{obs}$ (km s$^{-1}$)	& 110$\pm$5		& 90$\pm$9       		\\
   $(vsin~i)_{calc.}$ (km s$^{-1}$)	& 112.8$\pm$0.3		& 84.5$\pm$0.4		       	\\
   
   $J$, $H$, $K_s$ (mag)$^{*}$		& \multicolumn{2}{c}{7.697$\pm$0.035, 7.718$\pm$0.026, 7.757$\pm$0.020}	\\
$\mu_\alpha cos\delta$, $\mu_\delta$(mas yr$^{-1}$)$^{**}$ & \multicolumn{2}{c}{2.07$\pm$0.62, 0.81$\pm$0.53} \\
%$U, V, W$ (km s$^{-1}$)  & \multicolumn{2}{c}{4.77$\pm$0.70, 1.07$\pm$0.34, -6.82$\pm$0.80}\\ 
\hline  
  \end{tabular}
  \end{center}  
\medskip
{\rm *{\em 2MASS} All-Sky Point Source Catalogue (Cutri et al. 2003)} \\ 
{\rm **Newly Reduced Hipparcos Catalogue (van Leeuwen 2007)} \\ 
\end{table}

\section{Discussion and conclusion}
Since the sum-of-squares in the analysis of CS light curve is too large when compared to the DM's light curves, and there 
is a doubt on the light constancy of their primary comparison we take weighted mean orbital parameters obtained by the 
analysis of the DM's B and V light curves.The mean parameters obtained from the light curve analysis are: $i=73^{\circ}$.05$\pm$0.02,   
$r_1$=0.3291$\pm$0.0013, $r_2$=0.2464$\pm$0.0012, T$_{2}$=6\,200$\pm$200 K. The fractional radius of the secondary component exceeds its corresponding Roche lobe radius by about 4\%. Combining the results obtained from RVs analysis 
we have derived the astrophysical parameters of the components and other properties listed in Table\,4. The mass and radius of 
the massive primary star are derived with an accuracy of 0.6 and 1.4 per cent while for the less massive donor star 0.4 and 
0.5 per cent. The observed and computed rotational velocities of the components are in good agreement, showing nearly synchronize rotation. In Fig.\,5, we plot the location of V1898\,Cyg stellar components in $\log~T_{eff}-\log~L/L_{\odot}$ diagram. The 
evolutionary tracks for masses 6 and 1.16 M$_{\odot}$ are also shown in this figure. For constructing the solar metallicity 
evolutionary tracks, we used the Cambridge version of the {\sc stars code} which was originally developed by Eggleton (1971) 
and substantially updated by Eldridge \& Tout (2004). The continuous and dashed lines from left to right corner show the zero-age 
and terminal-age main-sequences, respectively. While the primary star locates very close to the ZAMS the secondary, less massive 
star appears to be evolved up to the giant branch as is common in semi-detached Algol-type binaries. This appearance of the 
components in the HR diagram is common for the classical Algol-type binaries. The donor seems to higher temperature and luminosity, in 
which most of the atmospheric material has been transferred to its companion. Since the secondary component fills its Roche lobe it transfers its mass to the more massive component. Therefore the orbital period change is attributed to the mass transfer. However, the gainer is ascended up to higher effective 
temperature and luminosity. In addition, V1898\,Cyg locates in the diagram between the specific angular momentum and mass-ratio 
where angular momentum decreases faster (see \,{I}banoglu et al. 2006).  The average distance to the system was calculated to be 
501$\pm$5 pc. However, the average distance to the system is estimated to be 621$^{+443}_{-182}$ pc from the trigonometric 
parallax measured by the Hipparcos mission. The distance derived in this study is smaller by about 20 per cent, the most 
importantly it has very small uncertainty when compared with that measured by Hipparcos mission. 

The North America (NGC 7000) and Pelican (IC 5070) nebulae (NAP) are known as the most nearby huge, extended HII regions where 
star formation with intermediate mass is still ongoing. The distance to this extended star-forming region has been estimated 
from 200 to 2000 pc (see for example, Bally and Reipurth, 2003). While Herbig (1958) estimates a distance of 500 pc, Laugalys 
and Strai{\v z}ys (2002) derived a distance to NAP 600$\pm$50 pc. If the star is a member of NAP complex the distance to the 
star derived by us is in a better agreement with that of Herbig, but agrees with that proposed by Laugalys and Strai{\v z}ys 
(2002) within $2\sigma$ level. Recently, Strai{\v z}ys, Corbally and Laugalys (2008) listed OB stars in the vicinity of 
NAP, one of which is V1898\,Cyg. The distance to the star estimated by us appears to confirm the membership of NAN. Proper 
motions for V1898\,Cyg are given in the SIMBAD database as $\mu_\alpha cos\delta$=2.07$\pm$0.62 mas yr$^{-1}$ and  $\mu_\delta$=0.81$\pm$0.53 
mas yr$^{-1}$ with a space velocity of 45.3 km s$^{-1}$. We selected about 20 stars in the vicinity of V1898\,Cyg listed by 
Laugalys and Strai{\v z}ys (2002) for comparison their mean proper motions with the variable. The velocities of the stars 
vary from 45 to $-97$ kms$^{-1}$, and the mean proper motions are: $\mu_\alpha cos\delta=0.48$, $\mu_\delta =-2.85$ mas yr$^{-1}$. It 
seems that one can not definitely classify which stars are actually belong to the NAP and which are not.        

\begin{figure}
\begin{center}
\includegraphics[width=7cm]{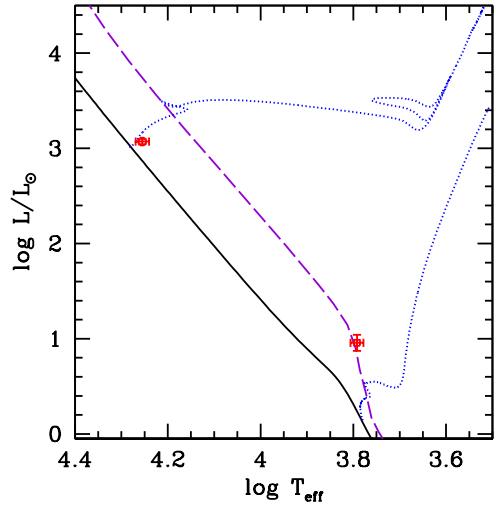}
\end{center}\caption{Location of the two stellar components of V1898\,Cyg in $\log~T_{eff}$-$~\log~L$ diagram, together 
with evolutionary models for the masses of 1.15 and 6.0 M$_{\odot}$. The open circle corresponds to the primary and the 
open square to the secondary with error bars.The zero-age main-sequence(continuous line) and terminal-age MS (long-dashed-dotted 
line)are also plotted. The evolutionary tracks are shown by dotted lines.}
\label{evrim;fig6}
\end{figure}

\section*{Acknowledgments}
We thank Prof.\ G.\ Strazzulla, director of the Catania Astrophysical Observatory, and Dr. G.\ Leto, responsible 
for the M. G. Fracastoro observing station for their warm hospitality and allowance of telescope time for the 
observations. We also thank to T\"UB\.ITAK National Observatory (TUG) for a partial support in using 
RTT150 with project numbers 10ARTT150-483-0 and 09BRTT150-468-0. This research has been also  
supported by T\"UB\.ITAK under project number 109T708 and INAF and Italian MIUR. This research has been made use of the ADS and CDS databases, operated 
at the CDS, Strasbourg, France and ULAKB{\.I}M S\"{u}reli Yay{\i}nlar Katalo\v{g}u. The authors thank to the anonymous referee for his/her valuable comments.

\end{document}